\documentclass[10pt, conference,compsocconf]{IEEEtran}
\usepackage{times}
\usepackage{subcaption}
\usepackage[font={small,it}]{caption}
\usepackage{url}
\usepackage{wrapfig}
\usepackage{cite}
\usepackage{tikz}
\usepackage{amsmath}

\usepackage{filecontents}
\usepackage{titling}
\date{}
\newcommand*\circled[1]{\tikz[baseline=(char.base)]{
          \node[shape=circle,draw,fill=black,text=white,font=\bf,inner sep=0.5pt] (char)
            {\scriptsize#1};}}
\usepackage{soul} 

\begin{document}

\pagenumbering{arabic}
\pagestyle{plain}

\title{\Large \bf Towards Designing a Self-Managed Machine Learning Inference Serving System in Public Cloud}
\author{\small{Jashwant Raj Gunasekaran, Prashanth Thinakaran, Cyan Subhra Mishra}, \small{Mahmut Taylan Kandemir, Chita R. Das}\\
 \small {\em  Computer Science and Engineering, The Pennsylvania State University, University Park, PA \quad} \\ 
}
\maketitle




\def\footnoterule{\kern-3pt
  \hrule \kern 2.6pt} 

\begin{abstract}
We are witnessing an increasing trend towards using Machine Learning (ML) based prediction systems, spanning across different application domains, including product recommendation systems, personal assistant devices, facial recognition, etc. These applications typically have diverse requirements in terms of accuracy and response latency, that have a direct impact on the cost of deploying them in a public cloud. Furthermore, the deployment cost also depends on the type of resources being procured, which by themselves are heterogeneous in terms of provisioning latencies and billing complexity. Thus, it is strenuous for an inference serving system to choose from this confounding array of resource types and model types to provide low-latency and cost-effective inferences. In this work we quantitatively characterize the cost, accuracy and latency implications of hosting ML inferences on different public cloud resource offerings. In addition, we comprehensively evaluate prior work which tries to achieve cost-effective prediction-serving. Our evaluation shows that, prior work does not solve the problem from both dimensions of model and resource heterogeneity. Hence, we argue that to address this problem, we need to holistically solve the issues that arise when trying to combine both model and resource heterogeneity towards optimizing for application constraints. Towards this, we envision developing a self-managed inference serving system, which can optimize the application requirements based on public cloud resource characteristics. In order to solve this complex optimization problem, we explore the high level design of a reinforcement-learning based system that can efficiently adapt to the changing needs of the system at scale.
\end{abstract}

\section{Introduction} 
\label{sec:intro}
\begin{figure}
\begin{minipage}[t]{0.95\linewidth}
\begin{center}
\includegraphics[width=0.9\textwidth]{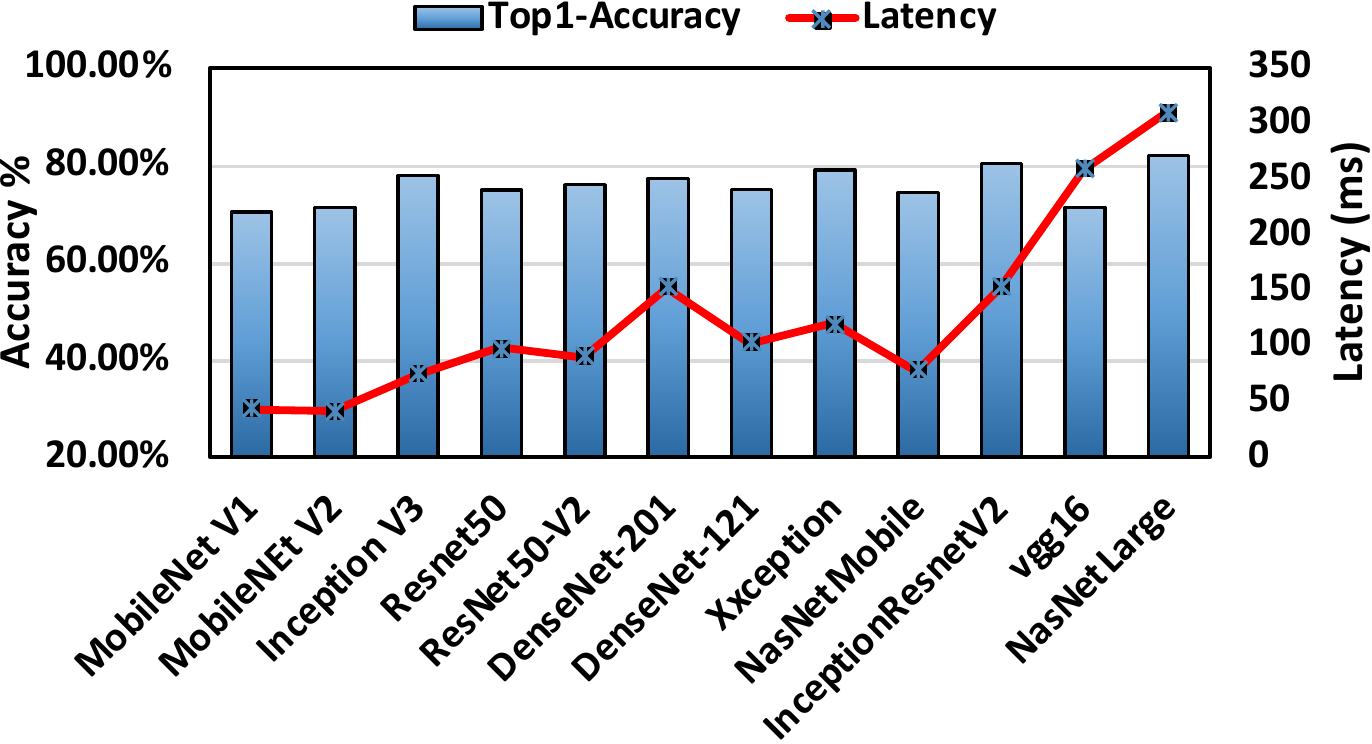}
\hspace{2mm}
\end{center}
\end{minipage}
\caption{Accuracy and Latency of Different ML Image Inference Models.}
\label{fig:models}
\end{figure}

Sustained advances in ML has fueled the proliferation of emerging applications such as product recommendation systems~\cite{facebook}, facial recognition systems~\cite{bartlett2005recognizing}, and intelligent personal assistants~\cite{sirius}. Among many ML paradigms, Deep Neural Networks (DNNs)~\cite{DNN}, owing to their generalization and massively-parallel nature, has been predominant in making all these applications pervasive and accessible to developers. A typical DNN model has two different phases, namely, \textit{training} and \textit{inference}. Training a DNN, which is the process of extracting and learning the patterns and the features from millions of sample-data, typically takes a few hours to days. 
The trained models can then be used to perform inferences, i.e., the classification task. Since typical large scale DNNs have millions of parameters and perform billions of multiplications and accumulations for executing a single inference, they are typically hosted as web-services, which are often queried for predictions. 
Conventionally, training is much more compute intensive~\cite{narayanan2019pipedream} (compared to an inference), takes many iterations and hence has been given considerable attention for better accuracy and convergence time. However, given the prevalence and demand of inferences, serving them on public cloud with a  tight bound of latency, throughput and cost is becoming increasingly more challenging~\cite{facebook}. 
\begin{figure}
\centering
\begin{minipage}{0.99\linewidth}
\centering
\includegraphics[width=0.9\textwidth]{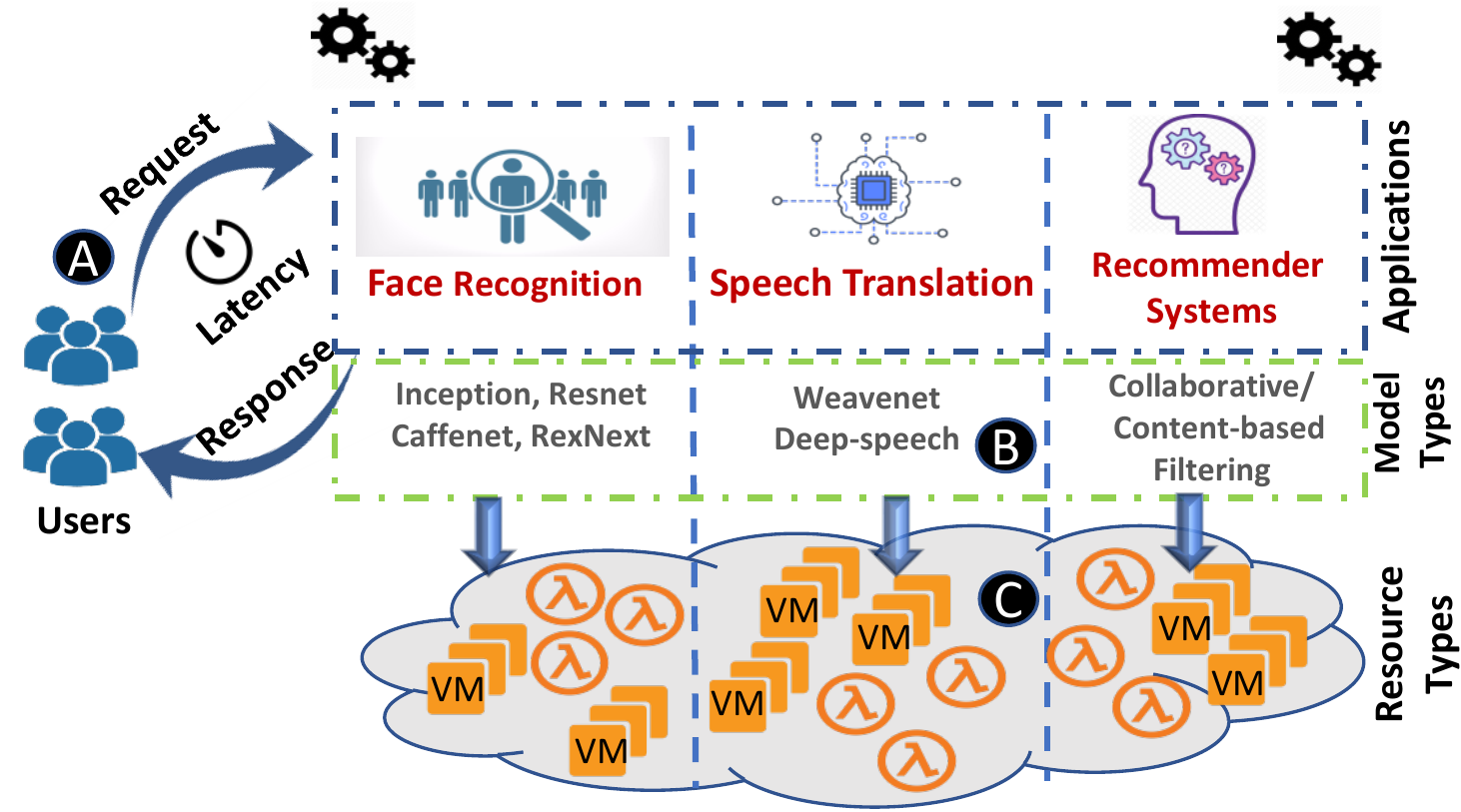}
\caption{Typical architecture of hosting a ML prediction-based system in public cloud.}
\label{fig:architecture}
\end{minipage}
\end{figure}
These inference queries are typically administered with strict response latencies of under one second~\cite{swayam} (Figure~\ref{fig:architecture}\circled{A}). 
Based on the application needs, prediction queries require different compute resources~\cite{8327042}, and have different accuracy, latency, and cost requirements. 
To ensure a required accuracy with given latency, applications have to  choose from a confounding array of different types of models (shown in Figure~\ref{fig:models}). 
Therefore, it is \emph{non-trivial for an application to choose the right model that can optimize for all requirements together}.  

Unlike accuracy and latency, which depends on the right model (Figure~\ref{fig:architecture}\circled{B}), the cost is dictated by the type of deployment used to host them in a public cloud  (Figure~\ref{fig:architecture}\circled{C}). The deployment costs differ based on the provisioning times and longevity of the resource procured. Typically these inference serving systems are hosted using Virtual Machines (VMs), which take a few minutes to start-up. Due to high start-up latencies~\cite{vm_startup}, using VMs for hosting ML services can lead to over-provisioning, especially during periods of poor workload predictability (flash crowds)~\cite{Urgaonkar:2008:ADP:1342171.1342172,Urgaonkar:2002:ROA:844128.844151}. In stark contrast to VMs, serverless functions have been made available by cloud providers, which can spin-up within a few seconds~\cite{firecracker}. As we will discuss in this paper, the cost of using VMs vs. serverless functions highly depends on the dynamically varying needs of the end-user query submission rates. 

Besides workload arrival rates, there is further variability in terms of configuring serverless functions to meet the end-user demands of latency and cost requirements. This is because, serverless functions are billed based on of number of invocations, compute time and memory requirement of the function. As we note in this paper, {the memory allocated per function is inversely proportional to compute time and directly proportional to billing cost}. 

Therefore, these apparent deficiencies of choosing the right resource type and selecting the appropriate model type for a given user requirement motivates the central question of this work: \emph{Does there exist an optimal resource procurement system which can balance the goals of diverse user requirements for accuracy, latency and cost, by efficiently mapping model parameters to heterogeneous resource specifications?} Our preliminary results suggest that using a combination of VMs and serverless functions could potentially provide a solution to this problem. 
As opposed to prior works~\cite{mark,spock}, which try to combine serverless functions with VMs to hide the start-up latencies of VMs, our primary interest lies in exploring the different \textbf{key aspects} to address when hosting DNN-based ML prediction serving systems in public cloud, as given below:

$\bullet$ \textbf{Diverse Models:} How to make the users oblivious of model selection from the extensive pool of models, for satisfying the accuracy, and latency requirements?  
     
$\bullet$ \textbf{Heterogeneous Public Cloud Resources:} What are the different options available in terms of combining different VM-based cloud services and serverless functions for a given user requirement?
    
$\bullet$ \textbf{Configuring Resources:} From the diverse options, how to right-size VMs and appropriately configure the serverless functions to efficiently cater to user specified cost, accuracy and latency constraint? 

$\bullet$ \textbf{Bring in Tune}: Based on the dynamically changing query arrivals over time, what is the right way to combine model diversity along with resource heterogeneity without compromising the user-specified requirements?

By exploring these key aspects, we envision developing a self-managed inference-serving system, which can provide for different diverse needs of applications by leveraging the heterogeneous resource availability from the public cloud. Towards this, we make the following \textbf{key contributions}. 

$\bullet$ We comprehensively characterize the cost, accuracy and latency implications of hosting ML inferences on different public cloud resource offerings and unravel the suitable model/resource configurations to meet the cost, latency and accuracy demands. 

$\bullet$ We quantitatively evaluate prior works~\cite{spock,mark,infaas} which are geared towards achieving this vision and show that they still suffer from several issues when trying to solve the complex problem of combining model and resource heterogeneity. 

$\bullet$ We propose detailed design choices that can adopted towards designing a self-managed inference-serving system. In addition, we design a scheme named Paragon on top of AWS platform, which incorporates some of the proposed design choices. Our initial results shows that Paragon can reduce cost of hosting ML prediction serving by up to 20\% when compared to the state-of-the-art prior works, for a given diverse accuracy and latency constraints.   

$\bullet$ Finally, we identify that it is still sub-optimal and time-consuming to design schemes for dynamically changing application requirements and resource configurations. To holistically solve this complex optimization problem by addressing all the challenges, we motivate the need for developing a reinforcement learning-based system, which can dynamically adapt to changing application needs and public cloud resource offerings.







\section{Characterization and Motivation} 
\label{sec:background}
\begin{figure}
\begin{minipage}[t]{0.95\linewidth}
\begin{center}
\begin{subfigure}[t]{.45\textwidth}
\includegraphics[width=0.99\textwidth]{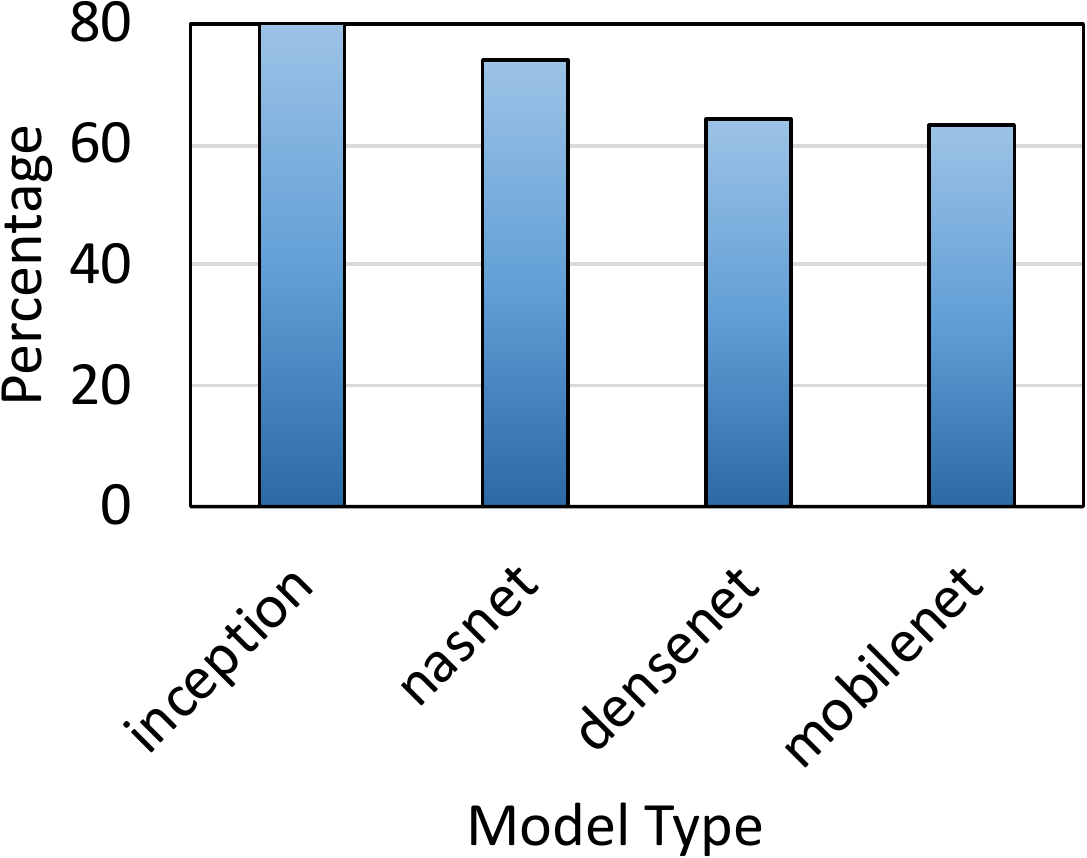}
\caption{Different accuracy for ISO-latency.}
\label{fig:iso-latency-accuracy}
\end{subfigure}
\hspace{2mm}
\begin{subfigure}[t]{0.45\textwidth}

\includegraphics[width=0.99\textwidth]{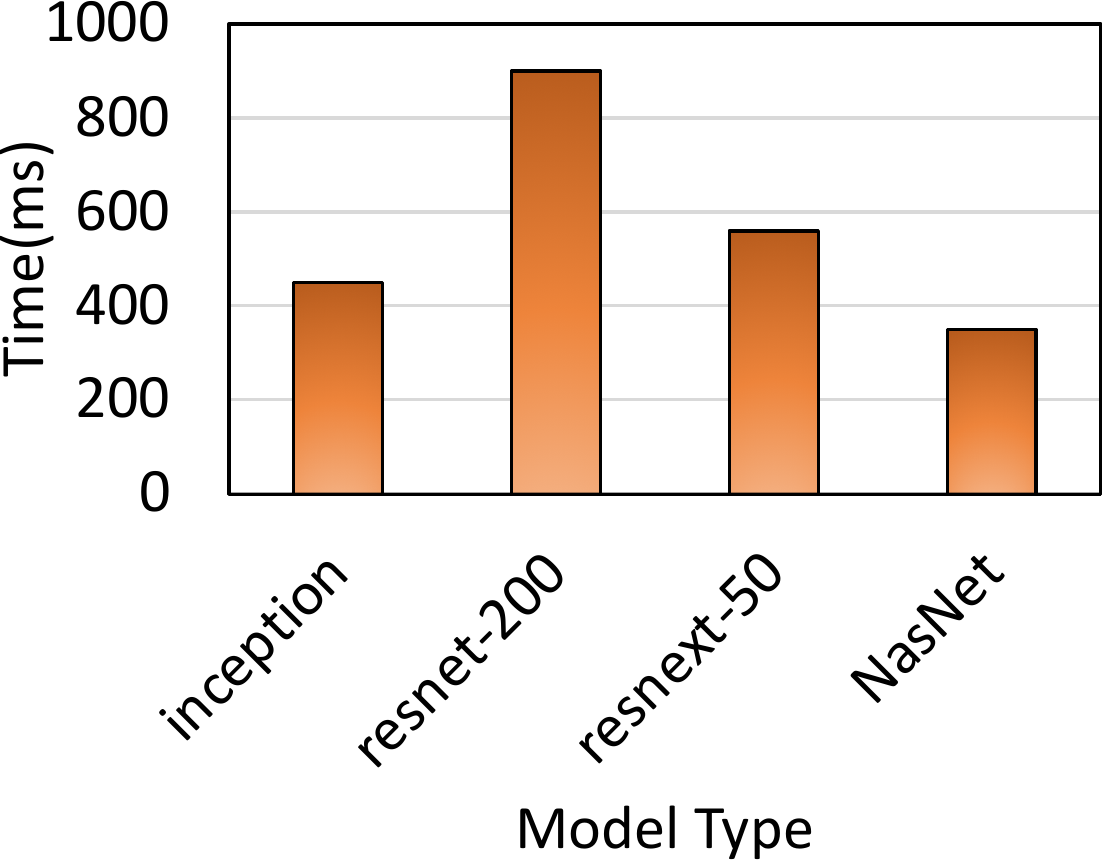}
\caption{Different response latencies for ISO-accuracy.}
\label{fig:iso-accuracy-latency}
\end{subfigure}
\end{center}
\end{minipage}
\caption{Comparison of different models under ISO-latency and ISO-accuracy setup.}
\label{fig:iso}
\end{figure}
\begin{figure*}
\centering
\begin{subfigure}[t]{.49\textwidth}
\centering
\includegraphics[width=0.8\textwidth]{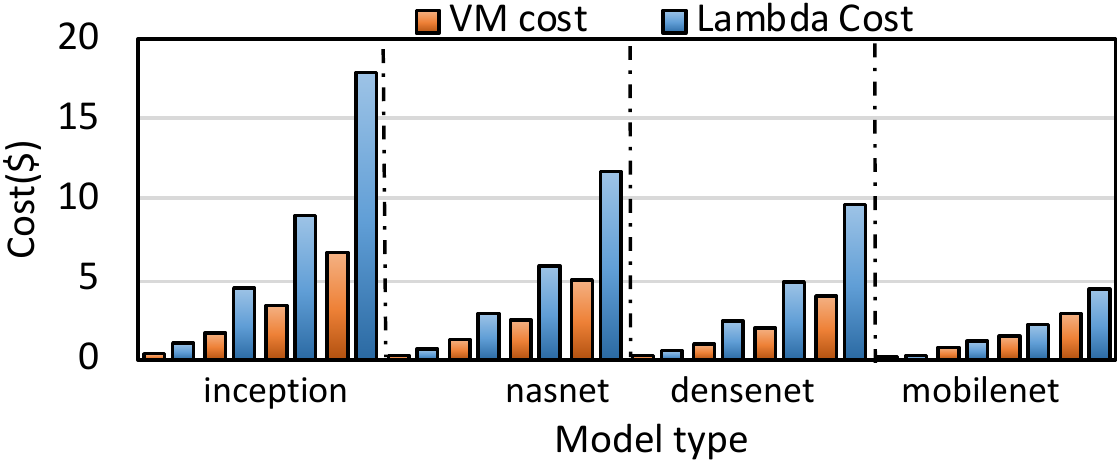}
\caption{Cost for ISO-latency models.}
\label{fig:iso-latency-cost}
\end{subfigure}
\centering
\begin{subfigure}[t]{0.49\textwidth}
\centering
\includegraphics[width=0.8\textwidth]{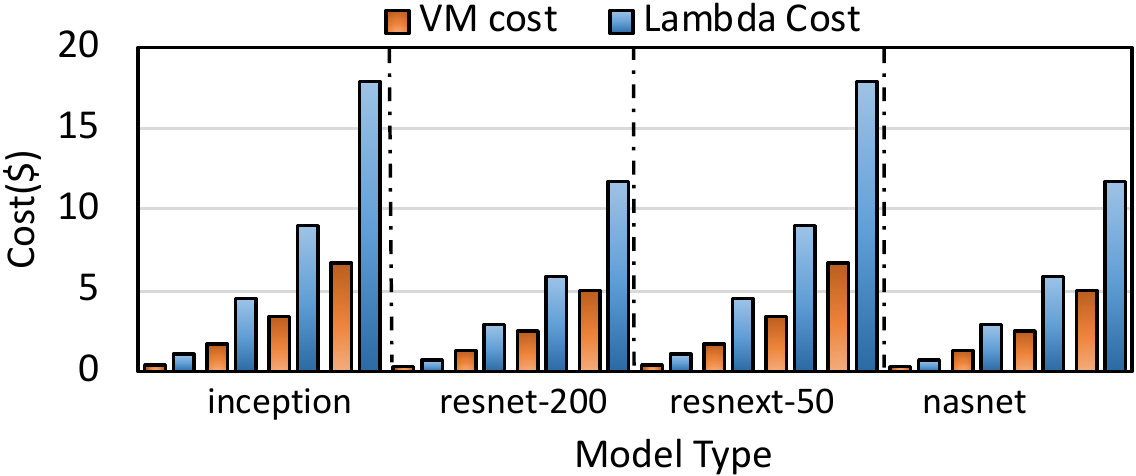}
\caption{Cost for ISO-accuracy models.}
\label{fig:iso-accuracy-cost}
\end{subfigure}
\caption{Variation of cost of using VMs vs. \emph{serverless functions} under constant request load. Each of the four bars under any model type corresponds to request arrival rates of 10, 50, 100, and 200 requests/second.} 
\end{figure*}
\subsection{Variability across model types}
Depending upon the accuracy and latency requirements of an end-user application, multiple models (shown in Figure~\ref{fig:models}) might satisfy a given constraint. For example, consider a face-recognition application that demands a response latency of under 500ms (ISO-latency). As shown in Figure~\ref{fig:iso-latency-accuracy},  four different models can satisfy the response latency, but each model comes with a different prediction accuracy. Similarly, if the same application requires accuracy to be at-least 80\% (ISO-accuracy), as shown in Figure~\ref{fig:iso-accuracy-latency}, four different models with different response latencies can satisfy the accuracy. Therefore, depending on the cost budget of the application, one can choose among the different model types by {\em trading-off accuracy or response latency.}. Hence, it is evident that there is a large optimization space where different models can be selected based upon the needs of the applications. \textbf{Prior work~}\cite{infaas} has tried to solve model selection only from a throughput perspective where different sized batching of multiple inference queries together results in varied throughput.\\ \textbf{Observation 1:} \emph{Model selection should be focused on meeting the cost requirement of an application without compromising on the accuracy and/or latency constraint.}
\subsection{Performance under given constraint}
\label{sec:vm-performance}
Model selection is not an independent problem because the user-applications also have a cost constraint incurred as a result of procuring resources from the public cloud. We compare the cost of deploying the inference service on a group of virtual machines and \emph{serverless functions}. We use m4-large instances for VMs and we fix the number of inference queries each VM can handle in parallel, without violating response latencies based on our characterization on AWS EC2. The \emph{serverless functions} are configured according to the memory requirements of each model. Figure~\ref{fig:iso-latency-cost} plots the cost of hosting the iso-latency model types (shown in Figure~\ref{fig:iso-latency-accuracy}) for a constant request\footnote{We use the terms request and query interchangeably.} arrival rates of 10, 50, 100, 200 req/sec over 1 hour duration. It can be seen that virtual machines are always cheaper compared to using \emph{serverless functions} for all constant request rates. A similar trend is observed for the iso-accuracy model types, which is shown in Figure~\ref{fig:iso-accuracy-cost}. 

There is also a possibility to use bigger VMs, which can handle more concurrent requests compared to m4-large, thus minimizing the total number of VMs used. However, we observe that the pricing of EC2 VMs is a linear function of the VM size in terms of compute capacity and memory. Hence, bigger VMs would still incur similar costs as smaller VMs. \\ \textbf{Observation 2:} \emph{ VMs should be used to handle requests during constant arrival rates. Also, the number of concurrent requests which can be executed in VMs should be accurately determined to meet response latency.}
\begin{figure}
\begin{minipage}{0.99\linewidth}
\centering
\includegraphics[width=0.82\textwidth]{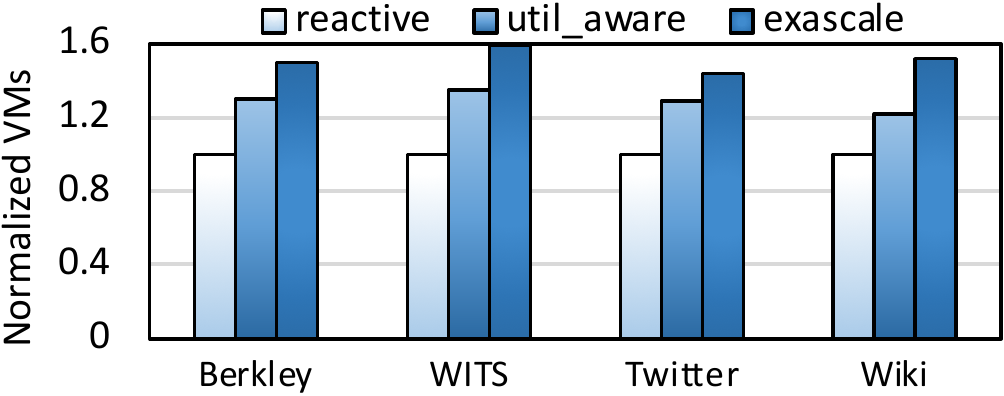}
\caption{Over-provisioning of \textit{util\_aware} and \textit{exascale}, normalized to a baseline \textit{reactive} scheme for four traces.} 
\label{fig:overprovision}
\end{minipage}
\end{figure}
\subsection{Over-provisioning VMs}
\label{sec:overprovision}
Real-world request arrivals rates are usually not constant as they significantly vary over time (e.g. diurnal, flash-crowds etc.) Therefore, resource procurement and management decisions need to be adjusted depending on the resource utilization/load and arrival rates. Public cloud providers leave these major decisions to be ``manually handled" by users, which is very time consuming and strenuous. As a result, the majority of application providers use {\em static resource provisioning}, which results in poor resource utilization and higher costs.

\textbf{Prior works}~\cite{infaas, hotspot, stratus} have tried to solve the resource scaling problem with respect to hosting the applications in virtual machines. They employ autoscaling mechanisms to cope up with respect to dynamic load variations. These autoscaling mechanisms can be of two types: (i) spawn VMs if the resource utilization of existing VMs reaches a certain threshold (80\% in most cases)~\cite{infaas}, and (ii) spawn additional VMs than predicted request demand~\cite{tributary}. We name the former autoscaling scheme as \textit{util\_aware} and the later as \textit{exascale}. Both these schemes suffer from over-provisioning VMs because (i) we cannot always accurately predict the future load, and (ii) resource utilization is not always the right indicator for increased load. 

\begin{figure}
\begin{minipage}{0.99\linewidth}
\centering
\includegraphics[width=0.82\textwidth]{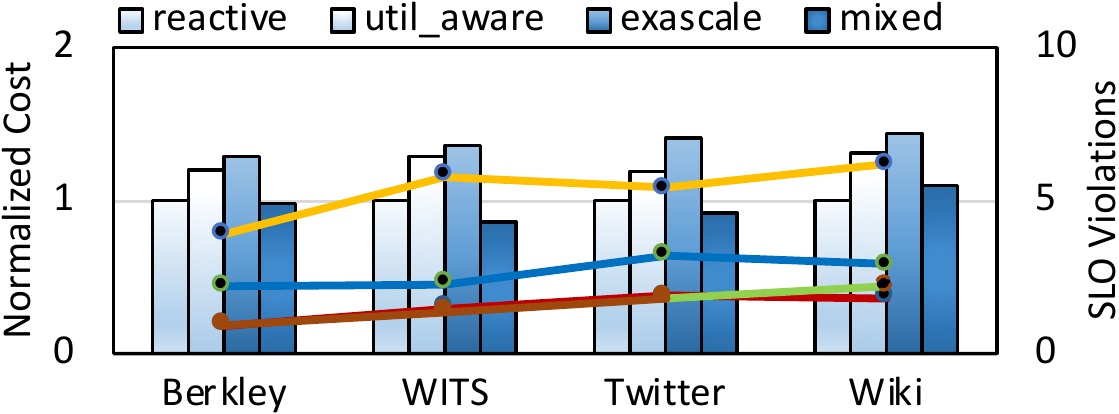}
\caption{Cost of using \emph{mixed} compared to  \textit{util\_aware} and \textit{exascale}, normalized to a baseline \textit{reactive} scheme for four traces. Percentage of SLA violations for each scheme are shown in the line graph (the corresponding color as in bar-graph is used for all the schemes.)}
\label{fig:mixed}
\end{minipage}
\end{figure}

We conduct simulation experiments to compare the schemes, using the profiled values (explained in Section~\ref{sec:vm-performance}) for four different well-known request arrival traces~\cite{berkley,wiki,twitter,wits}. Each request in the trace is associated with an ML inference query, which is randomly picked from our model pool. Figure~\ref{fig:overprovision} shows the ratio of over-provisioned VMs compared to a baseline \textit{reactive} autoscaling mechanism. It can be seen that although both \textit{util\_aware} and \textit{exascale} can reduce SLO violations (shown in Figure~\ref{fig:mixed}), they still suffer from 20\% to 30\% over-provisioned VMs across all four traces. This, in turn, increases the cost of deployment (shown in Figure~\ref{fig:mixed}), compared to baseline \textit{reactive} scheme. \\ \textbf{Observation 3:}\emph{ Only-VM based resource procurement should not be used during dynamic load as it leads to over-provisioned resources and increased cost.}
\begin{figure}
\begin{minipage}{0.9\linewidth}
\centering
\includegraphics[width=0.82\textwidth]{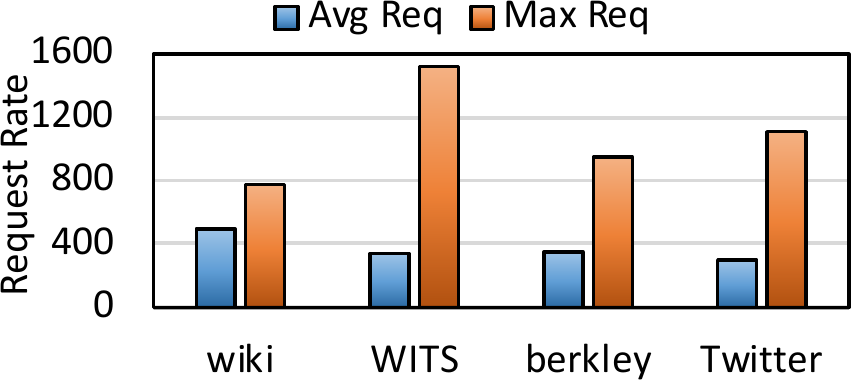}
\caption{Difference between Peak and median of request rates for four different traces.\vspace{-3mm}}
\label{fig:peak-median}
\end{minipage}
\end{figure}
\subsection{Using \emph{serverless functions} with VMs}
The provisioning latency is a major contributor for VM over-provisioning during request surges because the increased time to provision new VMs results in the increase of response latencies which in-turn leads to  provisioning more VMs in advance. \textbf{Prior works}~\cite{spock,mark} try to hide the provisioning latency of VMs by using \emph{serverless functions} as a handover mechanism when starting new VMs. We name this scheme as \emph{mixed} procurement. However, these schemes does not address the holistic problem by taking into account model selection, resource selection, and resource scaling to cope up with user-specified constraints. 
\begin{figure}[t]
\centering
\begin{minipage}{0.99\linewidth}
\begin{center}
\includegraphics[width=0.88\textwidth]{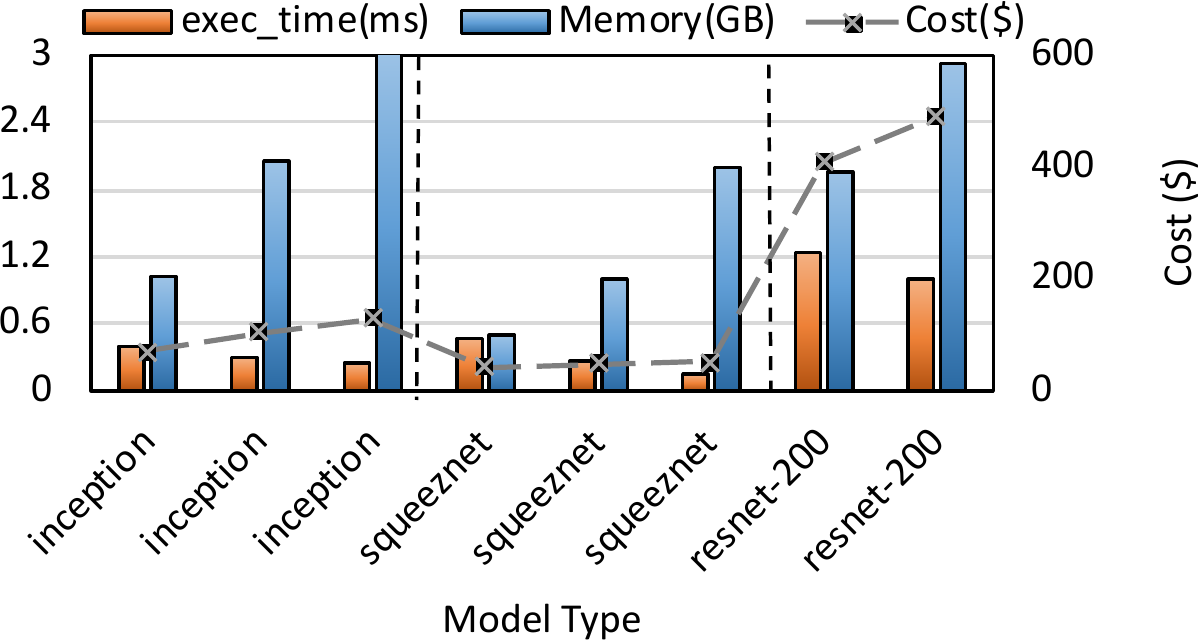}
\caption{Cost variation for different allocations in \emph{serverless functions}. Compute time (seconds) and Memory allocated (GB) is shown on left Y-axis. Cost (\$) is shown in right Y-axis.\vspace{-2mm}}
\label{fig:configuration}
\end{center}
\end{minipage}
\end{figure}
We conduct similar experiments to mimic the \emph{mixed} procurement scheme. As shown in Figure~\ref{fig:mixed}, \emph{mixed} procurement reduces the over-provisioning cost of VMs. At the same time it also  minimizes latency violations equivalent to \textit{exascale} scheme. However, we argue that there is scope to further optimize resource procurement based on the frequency of peak load and constant load in a given request arrival scenario. Figure~\ref{fig:peak-median} plots the peak-to-median ratio for three different traces. From our simulation experiments we observe that \textit{mixed} procurement did not reduce cost of Wiki trace. This is because the difference between peak-to-median of request traces is not large and therefore more functions get offloaded to \emph{serverless functions}.  Thus using \emph{serverless functions} for such scenarios will not drastically reduce cost. For the other traces like Berkeley, WITS and Twitter, the difference between peak-to-median is more than 50\% and therefore they can benefit from offloading requests to \emph{serverless functions}. \\
\textbf{Observation 4:}\textit{ It is important to note that, the request arrival pattern plays a key role in determining if \textit{mixed} procurement can be cost effective.}

\subsection{Challenges with \emph{serverless functions}}
Apart from arrival rates, memory allocation to \emph{serverless functions} play a non-trivial role in terms of cost. In our experiments we configure the memory allocation to the lambda function such that individual query latency is within the user-specified latency constraint. We conducted characterization experiments on AWS Lambda, currently the most predominant serverless function provider, to study the memory allocated vs computation time trade-off. Figure~\ref{fig:configuration} shows the computation time and cost for executing 1 million inference queries for three different model types with different memory allocations. We vary the memory allocation, starting from least required memory for the model to the maximum available limit in AWS (3GB)\footnote{For squeezenet model, allocating beyond 2GB did not reduce computation time, but resulted in increased cost}. It can be clearly seen that the computation time reduces with increased memory allocation but also results in higher cost of deployment for every model type. This is because, inherently, serverless providers allocate a powerful compute core for functions with higher memory allocation. Therefore, depending on the latency requirements of the user applications, \emph{serverless functions} need to be allocated the appropriate memory. However, this might result in increased cost when using \emph{serverless functions} along with VMs for varying latency requirements. Hence, the overall cost incurred by \emph{mixed} procurement can be higher or lower than VM-only autoscaling policies. 

\textbf{Prior work} like Cherrypick~\cite{cherrypick} solves the resource selection and configuration problems from VM perspective but does not consider Serverless Functions. We argue that compared to VMs there is more variability in configurations for \emph{serverless functions} because the resources are billed in a fine-grained allocation of CPU and memory.\\
\textbf{Observation 5:} \emph{Serverless functions can be used with VMs to avoid over-provisioning resources, but the right configuration needs to be accurately determined for the functions such that it satisfies the application cost and latency constraints.}

\section{How to Design Self-Managed ML Prediction Serving System?}
\label{sec:modeling}
The objectives from Section~\ref{sec:background} strongly motivate the need for a self-managed ML-prediction system that avoids the over-provisioning problem in VMs by efficiently blending \emph{serverless functions} with VMs. At the same time, right-sizing the number of requests in VMs and correctly configuring \emph{serverless functions} is quintessential to satisfy the three primary application constraints: cost, latency, and accuracy. 
\subsection{Model Selection}
\label{sec:model-selection}
In accordance with \emph{Observation 1}, model selection should be a function of any two parameters which optimize the remaining (third) parameter. Prior work ~\cite{infaas} solves an  optimization problem such that the input parameters are model\_type, hardware\_type (CPU or GPU), and the output parameter is response latency.  To do so, they suggest using offline profiling or results from previous executions. Unlike prior works, we suggest that the input and output parameters can be any linear combination of the three primary parameters mentioned above, depending on the application constraints. Note that, in contrast to cost and response-latency, accuracy cannot be determined just from the previous runs. We need some feedback from the end-user to make a correct estimate of accuracy. Therefore, it would be best to build a learning-based system, which takes into account feedback (user-given data) to build a novel model selection system. 
\subsection{Resource selection}
\subsubsection{Static Load} From \emph{Observation 2}, it is clear that,  besides model selection, it is crucial to select and configure the right resource to satisfy the application constraints. For applications where the request load is fairly constant over time, only VM-based resources can be procured to serve the requests. To determine the number of requests each VM can handle in parallel, we can conduct offline profiling for different model types.
\subsubsection{Dynamic Load} For applications with dynamic load (\emph{Observation 3}), \emph{serverless functions} can be used to mitigate the over-provisioning cost of VMs. However, a single application can contain a mix of queries with varying latency demands. Therefore, queries with strict latency requirements can be scheduled on \emph{serverless functions}, if a VM with free resources is unavailable. To handle dynamic load variations, a load-monitor can be designed such that it constantly monitors different periods of static load and peak load. We propose to plug-in intelligent peak-to-median prediction policies (in accordance to \emph{Observation 4}) , which can aid the load-monitor to estimate the duration of static load. Furthermore, it can measure the peak-to-median ratio in sampling windows, which can be used to decide if \emph{serverless functions} are required to balance the load. However, during flash-crowds, where load-prediction fails to accurately estimate the load, \emph{serverless functions} can inherently be used to handle requests to meet the response latency, but by incurring higher costs.
\subsubsection{Provisioning Time vs Execution Time} We know that new VMs take a few hundred seconds to start-up. \emph{Serverless functions} can start-up much faster (1s-10s), but they also incur additional latency to load a pre-trained model from external data-store. Prior literature~\cite{peeking,spock,mark} tries to hide the model load latency by pre-warming serverless function instances through periodically issuing dummy requests. However, such hacks can fail if the cloud service provider decides to change the idle timeout of function instances or change the overall mechanism to recycle idle function instances. Rather than capitalizing on such design hacks, we need to develop prediction policies to estimate load correctly. Also, we suggest service providers should handle the pre-warming decision by knowing model-wise usage statistics to enable instance sharing, which uses the same models. This would lead to a reduction in cold-start latencies incurred for users with the same type of requests.
\subsubsection{Configuring Serverless Functions}
In lieu with \emph{Observation 5}, it is quintessential to configure the memory allocation of \emph{serverless functions} to meet the application-specific response latency. Through offline profiling or initial runs, we can determine the right memory allocation for a given response latency. From our observations in \emph{AWS Lambda}~\cite{lambda}, three types of cores are allocated in the increasing order of the memory allocation (\textit{0.5GB, 1.5GB, and $>$2GB}). Also, these policies can be changed over time by Amazon, and they can also be different for other cloud providers~\cite{google,ibm, azure}. Therefore, the resource manager
should be able to leverage this information to make optimal serverless function configuration decisions.


\section{Diving into the Benefits of Self-Managed ML Serving System}
\label{sec:study}
\begin{figure*}
\begin{minipage}[t]{0.95\linewidth}
\begin{center}
\begin{subfigure}[t]{.32\textwidth}
\includegraphics[width=0.99\textwidth]{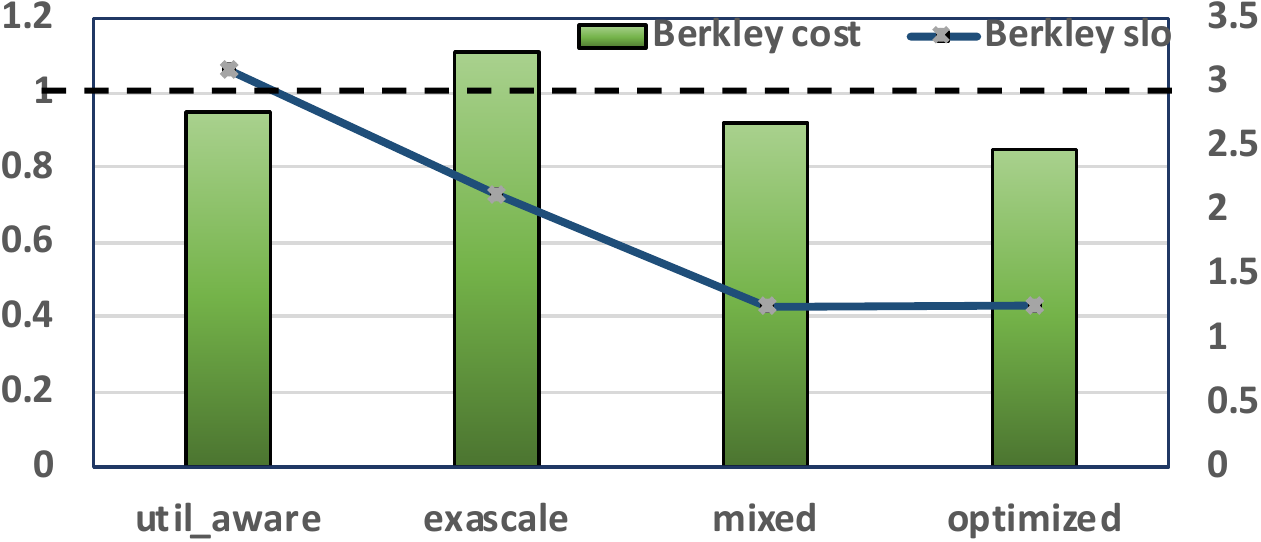}
\caption{Workload-1: Berkeley Trace.}
\label{fig:var-slo-b}
\end{subfigure}
\begin{subfigure}[t]{0.32\textwidth}
\includegraphics[width=0.99\textwidth]{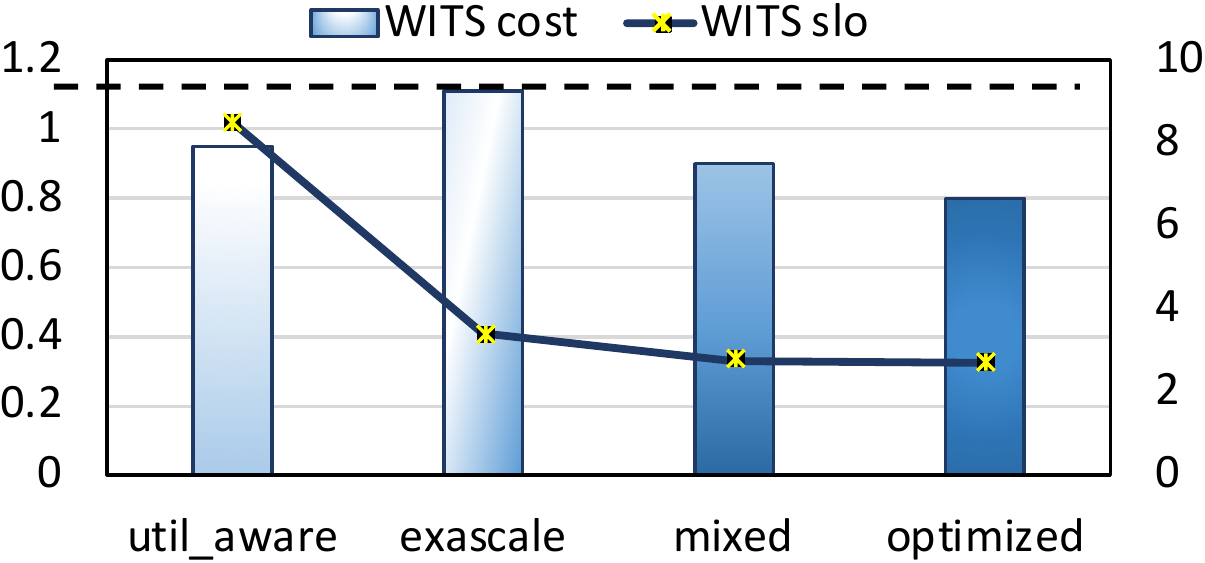}
\caption{Workload-1: WITS Trace.}
\label{fig:var-slo-w}
\end{subfigure}
\begin{subfigure}[t]{0.32\textwidth}
\includegraphics[width=0.99\textwidth]{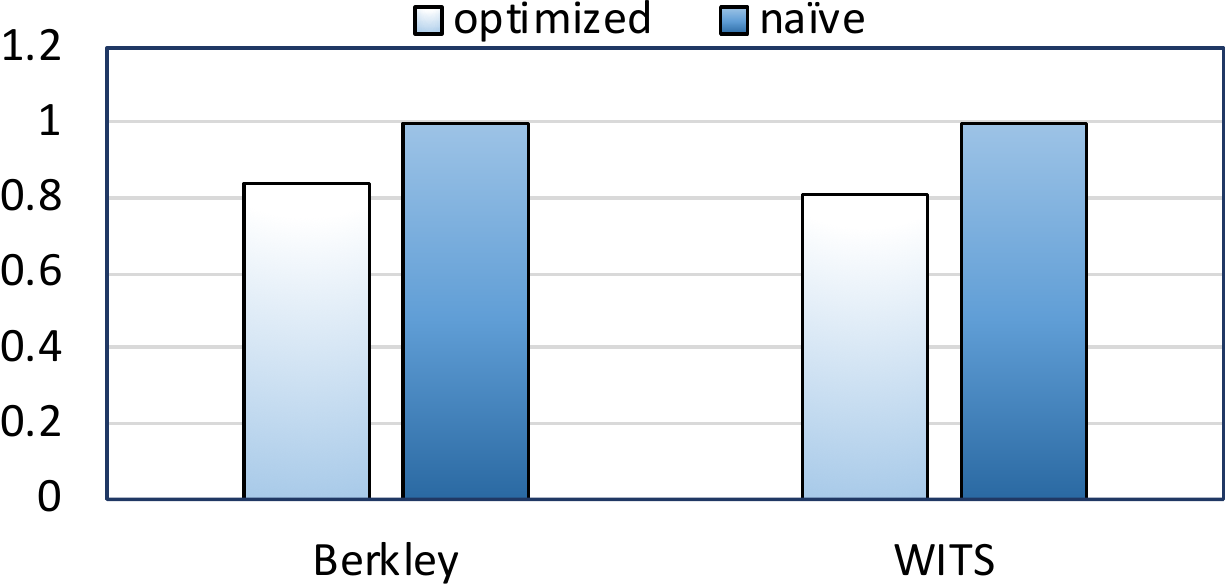}
\caption{Workload-2: Comparing Model Selection.}
\label{fig:var-const}
\end{subfigure}
\end{center}
\end{minipage}
\caption{Comparison of the resource procurement cost for two different traces using five different schemes. The primary y-axis is cost and secondary y-axis is slo violations. The results in Figure (a) and (b) are normalized to reactive scaling scheme. The results in (c) is normalized to naive model selection scheme.}
\label{fig:var-slo}
\end{figure*}
This section introduces how an ML-serving framework can capitalize on the design choices discussed in Section~\ref{sec:modeling}. We design three different experiments to study the effects on cost of ML servings due to (i) varying SLOs and (ii) varying application constraints.

\subsection{Implementation Methodology}
We developed a prototype on top of Amazon EC2
and Lambda services to evaluate the some of the benefits of our proposed design choices. We use AWS as the testbed for conducting extensive experiments. The types of instance used in our evaluation include all the c5 and m5 instances for EC2.  By offline profiling , we estimate the number of model instances each VM can execute in parallel without violating the model latency. Also, we estimate the right configuration of lambda functions by conduction offline experiments. 

For the model selection problem, we maintain an offline model cache which consists of the details of individual model latency and accuracy profiled by executing on c4\.large VM. The scheduler will pick the right model combinations from the cache based on the application requirements. We implement a load generator, which uses a 1 hour sample of the real-world trace for request arrival time generation. Each request is derived from a pool of pre-trained ML inference models for image classification (as explained in Section~\ref{sec:background}). We use \texttt{Mxnet}~\cite{mxnet} and \texttt{Tensorflow}~\cite{tensorflow} ML framework to deploy and run inference on the models.

\subsection{Evaluation}
We evaluate our results by comparing the cost, latency and accuracy for two different workloads. The first workload-type consists of a mix of queries which have both strict and relaxed latency requirements. We compare the execution of this workload against the following resource procurement schemes: (i) \textit{util\_aware}, (ii) \textit{exascale}, (iii) \textit{mixed} and (iv) \textit{Paragon}. These schemes are modeled after state-of-the-art prior works as explained earlier in Section~\ref{sec:overprovision}. The \textit{Paragon} scheme does not offload to lambdas for relaxed latency queries. 
The second workload-type consists of different cost, accuracy and latency requirements for all queries. We compare the \textit{Paragon} model selection scheme against a naive constraints-unaware model selection scheme. 
\subsection{Results}
In the follow section, we discuss the results of execution the two different workloads using Berkeley and WITS traces.
\subsubsection{Variable SLO}
Figure~\ref{fig:var-slo} plots the SLO and cost for workload-1 across Berkley and WITS trace. It can be seen that mixed scheme has similar cost to reactive but it reduces SLO violations by up to 60\%. This is because, the mixed scheme offloads request in the peak to serverless functions. However, the \textit{Paragon} scheme is 10\% more cost-effective than mixed and at the same time ensures similar SLOs. This is because the \textit{Paragon} scheme is aware of the latency requirements of individual queries and does not blindly offload queries to lambdas when there is increase in load. Therefore, this results in reduced cost and at the same time does not violate SLOs. This re-instantiates our claim that the resource procurement scheme needs to be aware of request constraints.
\subsubsection{Variable Constraints}
Figure~\ref{fig:models} shows the candidate models which can be used a given user of latency and accuracy. 
Our \textit{Paragon} scheme optimizes the model selection such that, it chooses the least cost-effective model for the given accuracy and latency constraint. The naive model selection policy would not choose the models as its oblivious to user requirements and model characteristics. From Figure~\ref{fig:var-const}, it can be seen that compared to naive selection scheme which does not optimize model selection for cost, the \textit{Paragon} schemes reduces the cost of resource procurement by up to 20\%. This is because the \textit{Paragon} scheme jointly considers all three parameters and chooses the least costing model.

\section{Towards Designing an Online Reinforcement learning-based System}
\label{sec:scheme}
All of the design choices we discuss in Section~\ref{sec:modeling} either majorly rely on offline profiling or results from initial runs. As mentioned earlier in Section~\ref{sec:model-selection}, these solutions can be time-consuming and are susceptible to failure if model types and resource types change over time. The results we discuss in the Section~\ref{sec:study} only show a sneak peak of the benefits which can be obtained but it does not holistically solve the self-managed prediction serving problem. Therefore the major challenge of provisioning a self-managed prediction serving in the cloud remains: \emph{how should a prediction serving system choose from this confounding array of resource types and model types to provide low-latency, cost-effective inference at scale?} To holistically address this problem at scale, the system has to be reconfigured in real-time, adapting to the changing system dynamics. Towards this, we discuss the high-level design of a policy gradient based-deep reinforcement learning approach~\cite{sutton2018reinforcement}, where the target policy is set, for example, as minimizing the overall cost or meeting response latency and/or accuracy constraint. Though we could use other standard supervised learning algorithms like regression~\cite{regression}, naive Bayes~\cite{naive-Bayes}, support vector machines (SVM)~\cite{svm}, etc., they cannot strike a balance between the policy exploitation and exploration at the same time since their reward functions (output parameters) are purely exploitative\footnote{Exploitation is limited to the scope of current policy -- by "exploration", we mean comparing existing policies to propose a new policy.}. This motivates us to develop a deep reinforcement learning based system that can jointly optimize for both both policy exploitation and exploration. 

Many existing works~\cite{mao2016resource,Mitra:2019:DLP:3343737.3343741} use a Monte-Carlo-based gradient policy~\cite{williams1992simple} suitable for predicting the request arrivals over time in traditional data-centers. However, our optimization problem is much more than just predicting the future request arrivals, rather we want to understand the execution results from current resource procurement  and model selection decisions and act on multiple reward policies (accuracy, response latency, or cost) rather than just one. Therefore, we plan to develop our policy based on multi-objective optimization using a deep reinforcement learning {ReML)} based approach. The objective function can be conservative in exploiting the known reward parameters such as cost and latency of individual applications so that it can optimize for overall system metrics such as QoS, query fairness and resource utilization. It is to be noted that ReML based approaches have been adopted by prior works such as Rafiki~\cite{rafiki} and Clipper~\cite{clipper} for selection policy when employing model-ensembling. These works have shown significant results using ReML, which supports our claim for applying ReML to solve the complex needs of a self-managed ML serving system.   

We plan to use proximal policy optimization (PPO)~\cite{schulman2015trust}, which adjusts its policy as long as it is within the desired policy target range by using an objective function. Therefore, cost reduction could be indirectly achieved by optimizing for other auxiliary objectives related to cost such as meeting the desired accuracy or response latency requirement.
\begin{figure}
\centering
\begin{minipage}{0.99\linewidth}
\begin{center}
\includegraphics[width=0.85\textwidth]{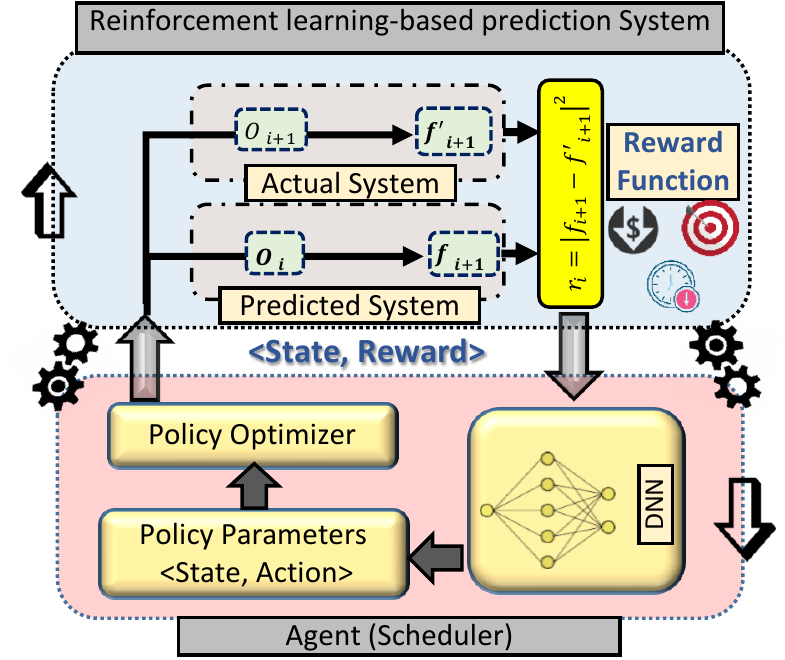}
\caption{Policy gradient based deep reinforcement learning approach.}
\label{fig:reml}
\end{center}
\end{minipage}
\end{figure}
The objective function with the PPO model is given as:
\begin{equation*}
L(\Theta )=\hat{E}_{t}[min(r_{t})(\Theta )\hat{A}_{t},r_{t}(\Theta),1-\varepsilon ,1+\varepsilon )\hat{A}_{t}] 
\label{eqn:ppo}
\end{equation*}
where, $\theta$ is the policy parameter, $\hat{E}_{t}$ denotes the expectations over every scheduling
decisions, $r_{t}$ is the ratio of probability under the new and old scheduling policies, $\hat{A}_{t}$ is the estimated policy improvement at time t and $\varepsilon$ is a hyper-parameter. $\varepsilon$ can be adjusted dynamically based on the degree of available resource and model types. 
Figure~\ref{fig:reml} shows a deep reinforcement learning-based system, which consists of two major components, namely, the agent and the system. At every time interval, the agent observes the system state o$_{i}$ and, based on it, it takes action a$_{i}$ in order to reach the system state \textit{f'}$_{i+1}$ (predicted), but in reality it reaches the system state \textit{f}$_{i+1}$ (actual). This disparity between the expected and observed system states is used to compute a transition reward r$_{i}$. Based on the r$_{i}$ and the actual system state \textit{f}$_{i+1}$, the agent incrementally learns to take the appropriate successive action a$_{i+1}$. The agent's decision process improves through this feedback system until the desired system state is reached. We propose to develop an online cost estimator model, which can at runtime determine if cost stays within user specified budget for using a combination of (i) serverless function of given configuration, and (ii) already available VMs. The cost estimator model would take into dynamic variation in request load such as ratio between peak-load to average-load over a given time window.  


\section{Discussion and Future Work}
\label{sec:discussion}
\subsubsection{Applicability to other applications} We believe the heterogeneous model selection, and resource selection problem would also be applicable to other application domains like query-processing and data-analytics. For instance, approximated queries which run faster but give lower prediction accuracy are best-suited to be executed in serverless functions. In data-analytics, frameworks like Spark have serverless-function based offerings as well. We can combine VM and serverless versions of Spark-based on the importance and frequency of analytics queries.  
\subsubsection{Limitations}  In our characterization studies, we did not include spot and burstable instances for VMs. We plan to consider them as well to develop a holistic system with all types of resource offerings from public cloud.
\subsubsection{Open Problems} We believe the problem discussed in this work can be addressed from several other dimensions. One promising way to look into the problem is to combine the DNN optimization techniques, e.g., model approximation, reduced precision network, ensemble learning, federated learning, etc., with our approach to deliver the most efficient solutions to the users with respect to cost and latency without (or minimally) compromising accuracy. For instance, with reduced precision activations, we can significantly lower the latency, and hence the cost, with a minimal dip in accuracy (approximately 2\% to 3\%). One more such example could include an ensemble learning approach, where the results from many low-cost and low-latency models with relatively lower accuracy could be aggregated together to give much higher accuracy.  In such scenarios, \emph{spot instances} can be helpful in lowering the cost. 
In a nutshell, we would like to have a holistic view of all public cloud offerings to build an efficient framework to offer  extremely intelligent, adaptive, yet low latency and low-cost solutions. 

\section{Conclusion}
\label{sec:conclusion}
There is wide-spread prominence in the adoption of ML-based prediction systems spanning across a wide range of application domains. The critical challenge of deploying ML prediction serving applications in public cloud is to combine both model and resource heterogeneity towards optimizing for application constraints.  In this paper, we propose to build a self-managed ML prediction system, which can optimize the diverse application requirements based on characteristics of heterogeneous public cloud resource offerings. Towards this, we discuss the trade-offs of intermixing resources like serverless functions along with VMs and identify the key challenges associated with configuring these resources. We propose multiple key-policies to make resource procurement and provisioning; (i) latency aware, (ii) multi-dimensional SLO aware, and (iii) request load variation aware. These policies can be used for cost-effective prediction serving without compromising on latency and accuracy In addition, we discuss the high level design of a reinforcement learning-based policy, which can dynamically adapt to the changing application needs and public cloud offerings.

\bibliographystyle{ieeetr}
\bibliography{references}
\end{document}